# Integrated high frequency aluminum nitride optomechanical resonators

Chi Xiong, Xiankai Sun, King Y. Fong, and Hong X. Tang[1]

*Department of Electrical Engineering, Yale University, 15 Prospect St., New Haven, Connecticut*

*06511, USA*

Aluminum nitride (AlN) has been widely used in microeletromechanical resonators for its excellent electromechanical properties. Here we demonstrate the use of AlN as an optomechanical material that simultaneously offer low optical and mechanical loss. Integrated AlN microring resonators in the shape of suspended rings exhibit high optical quality factor ($Q$) with loaded $Q$ up to 125,000. Optomechanical transduction of the Brownian motion of a GHz contour mode yields a displacement sensitivity of $6.2\times10^{-18}$m/Hz$^{1/2}$ in ambient air.

[1] e-mail: hong.tang@yale.edu

Vibrating micro- and nanoelectromechanical system (MEMS/NEMS) resonators is a scalable, energy efficient, and low cost technology to meet the ever growing demand for radio-frequency (RF) filters and frequency reference elements. MEMS resonators based on piezoelectric materials provide large electromechanical coupling coefficients and superior design flexibility [1]. Owing to its excellent mechanical properties and integration capability, aluminum nitride (AlN) has emerged as one of the most promising material platforms in enabling high-performance electromechanical devices including surface acoustic wave (SAW) devices [2], thin-film bulk acoustic wave resonators (FBARs) [3], contour-mode resonators [4] and lamb wave resonators [5]. While super high frequency AlN contour-mode electromechanical resonators operating at frequencies over 4 GHz have been demonstrated [6], it is inevitable that the ultimate operation bandwidth of electrically sensed resonators will be limited by parasitic electrical coupling and impedance mismatch. On the other hand, optical transduction can, in principle, achieve almost unconstrained bandwidth and has already demonstrated superior sensitivity [7,8]. Integrating AlN resonators with photonic circuits could enable high-performance electro-opto-mechanical oscillator systems leveraging the strong electromechanical coupling of the piezoelectric material and the wide bandwidth offered by optical readout.

In this study, we design and fabricate a monolithically integrated contour-mode AlN mechanical ring resonator which simultaneously serves as an optical cavity for sensitive displacement readout. With an optimized fabrication process, we measure high optical quality factor (loaded $Q$ > 125,000) in our suspended AlN ring resonators. The strong optomechanical interaction allows us to resolve the thermomechanical motions of the ring contour modes with frequency over GHz at room temperature and ambient pressure.

Our AlN ring resonators are fabricated in 330-nm-thick *c*-axis-oriented AlN thin films sputtered on oxidized silicon wafers. Fig. 1(a) shows an optical micrograph of a suspended AlN ring resonator made adjacent to a straight coupling waveguide. As shown in the picture, the ring is supported by four 0.5-μm-wide spokes which are anchored by a central disk sitting on underlying oxide. During the release process, a portion of the coupling waveguide would also be undercut due to the isotropic nature of the sacrificial oxide etching. Here, we employ a two-step lithography and dry-etching technique [9] to improve the device yield and avoid the spurious mechanical modes introduced by the suspended coupling waveguides. During the first electron beam lithography, we define the photonic patterns using hydrogen silsesquioxane (HSQ) resist and time the subsequent chlorine-based dry etching to ensure a 70-nm-thick remaining AlN slab layer. The second *e*-beam lithography is then performed using ZEP520A resist to define a release window surrounding the ring resonator. In this way, the residual layer of HSQ left from the first etch is used as a mask to etch through the remaining AlN slab within the release window, while the 70-nm AlN slab outside the release window is preserved and serves as a natural mask for the wet etch of the oxide in buffered oxide etchant. The wet etching is timed so that the center disk is not fully undercut. The devices are finally dried in a critical point dryer to prevent stiction. Fig. 1(b) shows a scanning electron microscopic image of a fabricated device. The coupling waveguide (denoted in red) adjacent to the ring resonator (denoted in green) is physically fixed to the slab layer, thus providing a stable optical coupling.

To realize GHz mechanical resonances, we focus on the radial-contour modes of the ring resonators with a geometry similar to electrostatically actuated "hollow-disk" resonators [10]. These radial-contour modes involve expansion and contraction motions about the ring width and their resonance frequencies can be approximately specified by $f = \frac{m}{2W}\sqrt{\frac{E}{\rho}}$, $m = 1, 2, 3, \cdots$, where $W$ is the width of the ring, $E$ and $\rho$ are the Young's modulus ($E_{AlN}$ = 330 GPa) and the density ($\rho_{AlN}$ = 3330 kg/m$^3$) of the material, respectively. To design a target 1 GHz resonance frequency with m = 1, the ring width needs to be $W$ = 5 μm. Furthermore, in order to minimize the anchor loss to the substrate and maximize the mechanical $Q$, the spoke length $L_{sp}$, defined as the distance from the wheel center to the ring attaching point, should be an odd multiple of a quarter wavelength of the spoke's longitudinal mode: $L_{sp} = (2n-1)\frac{\lambda_a}{4} = \frac{2n-1}{4f}\sqrt{\frac{E}{\rho}}, \quad n = 1, 2, 3 ....$ In this study, we choose the spoke length to be $L_{sp} = 13\lambda_a/4 = 32.6$ μm, and the inner and outer radii are $R_i = L_{sp}$ = 32.6 μm and $R_o$ = 37.6 μm accordingly. This footprint of the ring will be sufficient to achieve a minimal waveguide bending loss for the optical cavity to yield a high optical $Q$.

In Fig 1(c), we plot the calculated frequencies for each radial-contour mode as a function of the outer ring radius $R_o$ when the inner radius of the ring is fixed at $R_i$ = 32.6 μm [11]. At $R_o$ = 37.6 μm (vertical dashed line), the theoretical values of the resonant frequencies are found to be 1.05 GHz (m = 1), 2.11 GHz (m = 2), etc. Also shown is the lower frequency 47 MHz "breathing" mode. Using the ring's optimized geometric parameters, we numerically simulate (COMSOL Multiphysics) the mechanical modes for the spokes-supported ring and obtain an excellent agreement with the theoretically predicted frequencies. In Fig 1(c) inset, we show the simulated displacement profile for the first two modes at 47 MHz ("breathing" mode) and 1 GHz (m = 1 "pinch" mode).

Light from a tunable diode laser source is coupled into and out of the device using a pair of grating couplers. A fiber polarization controller is used before the input grating coupler to select the TE-like polarization. The transmitted light is detected by a low-noise InGaAs photoreceiver (New Focus 2011). Fig 2(a) shows the transmission spectrum of a typical released ring resonator with low input power (-20 dBm before the input coupler). Because the width of the ring (5 μm) allows for multi-mode propagation, different sets of resonances show up with different extinction ratios. The free spectral range (FSR) is found to be 5.0 nm. Based on $FSR = \lambda_0^2/(2\pi n_g R)$, the group index is calculated to be $n_g$ = 2.0. Fig 2(b) shows the zoom-in spectrum of a near critically coupled resonance at 1544.85 nm with extinction ratio of 18 dB and a fitted $Q$ factor of 125,000.

To optically transduce the mechanical modes, we tune the input laser to the wavelength corresponding to the maximum slope of an optical resonance and record the RF power spectrum of the optical transmission. At a high input power and blue cavity detuning, the optomechanical back-action can overcome the mechanical damping and cause the sustained mechanical self-oscillation [12]. As shown in the inset of Fig. 3(a), above a threshold optical power of +15 dBm (before the input coupler), the ring resonator oscillates in the fundamental radial-breathing mode at 47.3 MHz, indicated by a series of harmonics of the fundamental mode recorded by the spectrum analyzer. To resolve the intrinsic mechanical $Q$ factors, we thus keep the laser input power low enough (+3 dBm) that the optomechanical back-action is negligible. Under this condition, the transmitted light is amplified by an erbium doped optical amplifier (Pritel FA-20) and then detected by a high-speed InGaAs photoreceiver with 1 GHz bandwidth (New Focus 1611). The output of the photoreceiver is analyzed by a spectrum analyzer. Figure 3(a) shows the RF spectrum of the mechanical modes of the wheel resonator measured in air. Three mechanical modes are found at 30.6 MHz, 47.3 MHz and 1.04 GHz.

Figure 3(b)-3(d) show zoom-in spectra of the resonances at respective frequencies. All the measurements are carried out at room temperature and atmospheric pressure. The 1st contour mode ("breathing" mode) is detected at 47.3 MHz with a measured $Q$ of 1,742 [Fig. 3(c)]. The 2nd contour mode ("pinch" mode) at 1.04 GHz has a slightly higher measured $Q$ of 2,473 [Fig.

3(d)]. This could be due to the fact that the 2$^{nd}$ radial-contour mode's mechanical motions are more confined on the ring. In addition, a 30.64-MHz resonance is observed and identified as one of the wineglass modes (circumferential number n = 4, "square" mode [13]) with a measured $Q$ of 1,200 [see the simulated displacement profile in Fig. 3(b) inset]. By comparing the expected displacement noise with the measured RF spectral density, we can calibrate the amplitude of Brownian motions for each mechanical mode. The displacement sensitivity then corresponds to the noise floor of the spectrum. (The spectral density of the thermomechanical displacement noise at resonance frequency is $S_x^{1/2} = \sqrt{4k_B TQ/m_{eff}(2\pi f)^3}$, where $k_B$ is the Boltzmann constant, $T$ the absolute temperature (300 K), $Q$ the mechanical quality factor, $m_{eff}$ the effective modal mass, and $f$ the resonance frequency. The effective modal mass is defined as $m_{eff} = \frac{m_0}{u_{max}^2 V} \iiint u(x,y,z)^2 dxdydz$, where m$_0$ is the physical mass of the ring, $u(x,y,z)$ is the displacement, $u_{max}$ is the maximum displacement, V is the volume of the ring.) Table I summarizes the simulated and experimental resonance frequencies, together with the simulated modal effective mass $m_{eff}$ and the calibrated displacement sensitivity $S_x^{1/2}$ of the three measured modes. The table shows that the experimental frequencies are in good agreement with their simulated values. It is also noteworthy that the 1.04 GHz mode shows the best displacement sensitivity of 6.2×10$^{-18}$ m/√Hz.

The GHz AlN optomechanical resonators demonstrated here can find applications in precision oscillators and high-speed ultrasensitive systems. One of the strategies to further enhance the optomechanical coupling is to employ rings with smaller radii, as the coupling strength $g_{om}$ = d$\omega$/dR is inversely proportional to the radius $R$. Second, even higher mechanical $Q$ can be possibly obtained by further mechanical optimization to minimize the radiation loss due to the supporting spokes [14].

In summary, simultaneous high optical $Q$ (125,000) and mechanical $Q$ (2,473) are observed in AlN optomechanical ring resonators vibrating at GHz frequencies. Compared with conventional materials for integrated cavity optomechanical systems (such as Si and $Si_3N_4$), AlN offers many unique optical and mechanical properties. One of the benefits is that AlN has a very large bandgap (6.2 eV), which provides superior suppression of two-photon absorption and thus enables more stable optical resonators with high power handling capability. More importantly, AlN has a large piezoelectric coefficient and an intrinsic electro-optic Pockels' effect [15], which make AlN resonators an interesting platform to implement tunable, electrically driven and optically sensed oscillator systems.

The authors acknowledge funding support from DARPA/MTO, the Packard Foundation and the NSF grant through MRSEC DMR 1119826. The authors thank Michael Power and Dr. Michael Rooks for assistance in device fabrication.

**References:**

1. R. B. Karabalin, M. H. Matheny, X. L. Feng, E. Defay, G. Le Rhun, C. Marcoux, S. Hentz, P. Andreucci, and M. L. Roukes, Appl Phys Lett **95** (10) (2009).
2. H. Okano, N. Tanaka, Y. Takahashi, T. Tanaka, K. Shibata, and S. Nakano, Appl Phys Lett **64** (2), 166 (1994).
3. C. M. Yang, K. Uehara, S. K. Kim, S. Kameda, H. Nakase, and K. Tsubouchi, presented at the Ultrasonics, 2003 IEEE Symposium on, 2003 (unpublished).
4. G. Piazza, P. J. Stephanou, and A. P. Pisano, J Microelectromech S **15** (6), 1406 (2006).
5. C. M. Lin, T. T. Yen, Y. J. Lai, V. V. Felmetsger, M. A. Hopcroft, J. H. Kuypers, and A. P. Pisano, IEEE T Ultrason Ferr **57** (3), 524 (2010).
6. M. Rinaldi, C. Zuniga, C. J. Zuo, and G. Piazza, IEEE T Ultrason Ferr **57** (1), 38 (2010).
7. M. Li, W. H. P. Pernice, and H. X. Tang, Nat Nanotechnol **4** (6), 377 (2009).
8. M. Eichenfield, J. Chan, R. M. Camacho, K. J. Vahala, and O. Painter, Nature **462** (7269), 78 (2009).
9. W. H. P. Pernice, C. Xiong, C. Schuck, and H. X. Tang, Appl Phys Lett **100** (9), 091105 (2012).
10. S. S. Li, Y. W. Lin, Y. Xie, Z. Y. Ren, and C. T. C. Nguyen, Proc IEEE Micr Elect, 821 (2004).
11. C. V. Stephenson, J Acoust Soc Am **28** (1), 51 (1956).
12. S. Tallur, S. Sridaran, and S. A. Bhave, Opt Express **19** (24), 24522 (2011).
13. Y. Xie, S. S. Li, Y. W. Lin, Z. Y. Ren, and C. T. C. Nguyen, IEEE T Ultrason Ferr **55** (4), 890 (2008).
14. X. K. Sun, K. Y. Fong, C. Xiong, W. H. P. Pernice, and H. X. Tang, Opt Express **19** (22), 22316 (2011).
15. P. Graupner, J. C. Pommier, A. Cachard, and J. L. Coutaz, J Appl Phys **71** (9), 4136 (1992).

**Figures:**

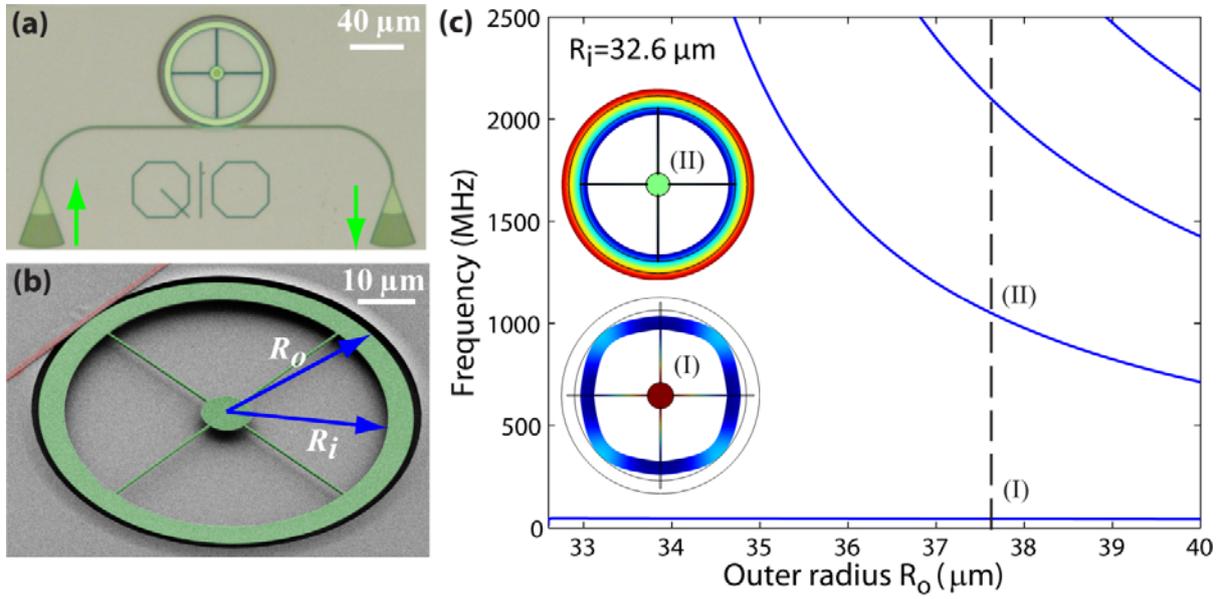

Figure 1 (a) Optical image of a fully integrated AlN optomechanical resonator. Light is coupled in and out of the chip using grating couplers. (b) False-color scanning electron micrograph of the fabricated AlN ring resonator (in green). The coupling waveguide is denoted in red. (c) Theoretical resonance frequencies of the radial-contour modes of the AlN ring resonators as a function of the outer radius of the ring. The inner radius of the ring is 32.6 µm. For an outer radius of 37.6 µm, the theory predicts a 1$^{st}$ order (I) radial-contour mode at 44.9 MHz and a 2$^{nd}$ order (II) radial-contour mode at 1.05 GHz. The insets show the numerically simulated displacement profiles for the two modes respectively (not to scale).

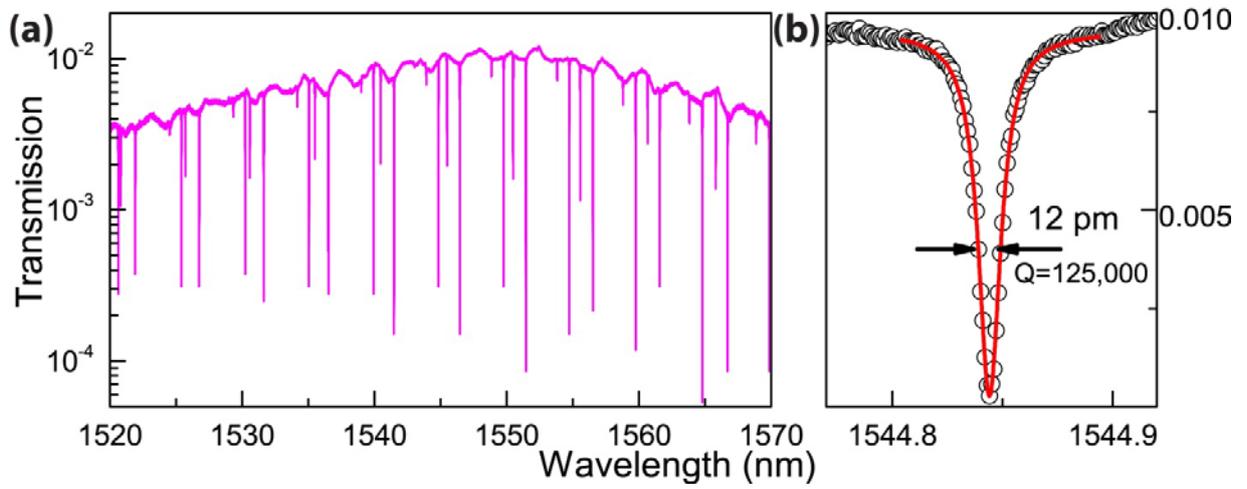

Figure 2 (a) Optical transmission spectrum of a released AlN optomechanical ring resonator measured in air with -20 dBm of input power before the input grating coupler. The free spectral range is 5.0 nm. (b) Zoom-in of one resonance at 1544.85 nm with a fitted linewidth of 12 pm, corresponding to an optical quality factor of 125,000. The optical resonance is near critically coupled with an extinction ratio of 18 dB.

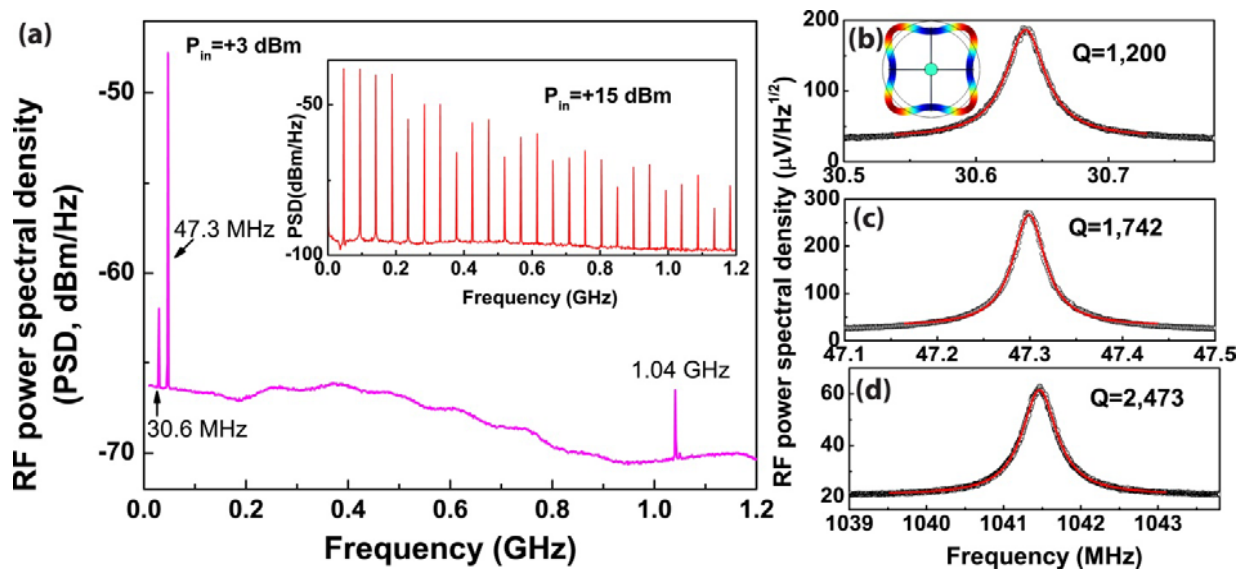

Figure 3 (a) Optically transduced RF spectrum of the mechanical modes of the AlN ring resonator measured in air with an input power of 3 dBm. Three mechanical modes are found at 30.6 MHz, 47.3 MHz, and 1.04 GHz. The inset is a measurement at +15 dBm optical power before the input grating coupler, showing backaction induced self-oscillations of the fundamental radial-contour mode (47.3 MHz). (b), (c), and (d) are the zoom-in spectra of the resonance at 30.6 MHz, 47.3 MHz, and 1041.5 MHz, respectively. The fitted (red lines) quality factors are 1,000, 1,742, and 2,473 respectively for the three mechanical modes measured in air. The inset in (b) is the numerically simulated displacement profile for the wine-glass mode (circumferential number n = 4) at 30.6 MHz.

TABLE I. Properties of the observed mechanical modes: simulated (superscript *s*), and experimental (superscript *e*) frequencies, effective modal mass ($m_{\text{eff}}$), measured quality factor ($Q_{\text{m}}$) and displacement sensitivity ($S_x^{1/2}$) of the ring resonator with inner radius $R_i$ = 32.6 μm and outer radius $R_o$ = 37.6 μm.

| Mode type | $f^{(s)}$ (MHz) | $f^{(e)}$ (MHz) | $m_{\text{eff}}^{(s)}$ (ng) | $Q_{\text{m}}^{(e)}$ (air) | $S_x^{1/2}$ (m/√Hz) |
|---|---|---|---|---|---|
| "square" wine-glass, 1st | 30 | 30.6 | 0.63 | 1200 | 3.5×10$^{-16}$ |
| radial-contour, 1st | 47 | 47.3 | 1.14 | 1742 | 9.3×10$^{-17}$ |
| radial-contour, 2nd | 1050 | 1041.5 | 0.42 | 2473 | 6.2×10$^{-18}$ |